\documentclass[aip,rsi,reprint,longbibliography]{revtex4-1}
\usepackage{graphicx}
\usepackage{subfigure}
\usepackage{units}
\usepackage[colorlinks=true,urlcolor=blue,linkcolor=blue,citecolor=blue,bookmarks=false,
breaklinks=true]{hyperref}
\usepackage{breakurl}

\begin{document}

\title{A compact linear accelerator based on a scalable microelectromechanical-system RF-structure}

\author{A.~Persaud}
\email[Author to whom correspondence should be addressed. Electronic mail: ]{apersaud@lbl.gov}
\author{Q.~Ji}
\author{E.~Feinberg}
\author{P.~A.~Seidl}
\author{W.~L.~Waldron}
\author{T.~Schenkel}
\affiliation{E.~O. Lawrence Berkeley National Laboratory, 1 Cyclotron Road, Berkeley, CA 94720, USA}

\author{A.~Lal}
\author{K.~B.~Vinayakumar}
\author{S.~Ardanuc}
\author{D.~A.~Hammer}
\affiliation{SonicMEMS Laboratory, Cornell University, Ithaca, NY 14853, USA}

\begin{abstract}
  A new approach for a compact radio-frequency (RF) accelerator
  structure is presented. The new accelerator architecture is based on
  the Multiple Electrostatic Quadrupole Array Linear Accelerator
  (MEQALAC) structure that was first developed in the 1980s. The
  MEQALAC utilized RF resonators producing the accelerating fields
  and providing for higher beam currents through parallel beamlets
  focused using arrays of electrostatic quadrupoles (ESQs). While the
  early work obtained ESQs with lateral dimensions on the order of a
  few centimeters, using printed circuits board (PCB), we reduce the
  characteristic dimension to the millimeter regime, while massively
  scaling up the potential number of parallel beamlets.  Using
  Microelectromechanical systems scalable fabrication
  approaches, we are working on further reducing the characteristic
  dimension to the sub-millimeter regime. The technology is based on
  RF-acceleration components and ESQs implemented in PCB or silicon
  wafers where each beamlet passes through beam apertures in the
  wafer. The complete accelerator is then assembled by stacking these
  wafers. This approach has the potential for fast and inexpensive
  batch fabrication of the components and flexibility in system design
  for application specific beam energies and currents. For prototyping
  the accelerator architecture, the components have been fabricated
  using PCB. In this paper, we present proof of concept results of the
  principal components using PCB: RF acceleration and ESQ
  focusing. Ongoing developments on implementing components in silicon
  and scaling of the accelerator technology to high currents and beam
  energies are discussed.
\end{abstract}

\maketitle

\section{Introduction}
Ion beams have many applications in research and industry. Increasing
intensity and reducing the size and cost are important aspects of
developing new particle accelerators for many possible applications.

Among them the most ambitious applications for high-intensity beams are
driving nuclear fusion power systems for the purpose of electricity
generation.  At present, there are two main approaches to achieving
practical fusion energy production, magnetic fusion energy (MFE) and
inertial fusion energy (IFE).  In both approaches, high intensity
particle beams have been proposed as methods to heat the fusion fuel.

In MFE, typified by toroidal devices called tokamaks, magnetically
confined deuterium and tritium plasma would be heated to fusion
conditions, i.e., an ion temperature of $\unit[\sim\!\!10]{keV}$, at
least in part, by intense \unit[0.1-1.0]{MeV} neutral deuterium beams
in order to initiate fusion reactions.  In principle, the resulting
charged reaction product, the \unit[3.5]{MeV} helium nucleus, will be
able to maintain the plasma temperature above \unit[10]{keV}
indefinitely in a well designed magnetic confinement system.  The
International Thermonuclear Experimental Reactor (ITER), now under
construction in France,\cite{Takatsu2011} will have two
\unit[16.5]{MW} deuterium beam systems with negative ions accelerated
to \unit[1]{MeV} kinetic energy before being neutralized and injected
into the device.  Each beamline occupies a volume exceeding
$\unit[15\times 5\times 5]{m^3}$.\cite{Hemsworth}

The idea of using ion beam accelerators for IFE arose in the 1970’s
shortly after the recognition that lasers might drive inertial fusion
targets.\cite{Nuckolls}  In laser and ion beam driven targets, the
fuel is a spherical shell a few millimeters in diameter.  Ion beam
energy of the order of \unit[1-10]{MJ} must be delivered in
$\unit[\sim\!\!10]{ns}$ to achieve fusion ignition, and computer simulations
indicate that $\sim\!\!100\times$ more energy may be created from the fusion
reactions before the compressed fuel disassembles.\cite{Bangerter}  In
contrast to the magnetic fusion approach, the process here is pulsed,
with a repetition rate of several Hertz.  The ion beam requirements
are constrained by the target design: for heavy ions, $A>100$, a
kinetic energy of several \unit{GeV} and currents in the kiloampere
range are required; for lighter ions and the same total beam energy,
since the \unit{MeV} per nucleon must be similar, the current must be
increased appropriately.

There is also interest in magneto-inertial fusion (MIF), where aspects
of magnetic and inertial fusion approaches are merged. Initially
low density deuterium-tritium plasma is confined by a magnetic field.
The plasma and embedded magnetic field are compressed by, e.g., a
metal liner, directed plasma or beams, with a confinement time longer
than characteristic of IFE and shorter than MFE.\cite{Lindemuth, Wurden2016}

For the IFE and MIF fusion energy applications, the required total
energy per pulse $>\!\!\unit[1]{MJ}$. For example, consider indirect
drive IFE driven with \unit[5]{MJ} pulses of \unit[5]{GeV} heavy ions
in a \unit[10]{ns} pulse.  The corresponding total beam current is
$\unit[100]{kA}$ and is distributed among approximately 100
beams. With present-day accelerator technology, the accelerators
generating the fusion driver beams would be kilometers long and are
usually the most costly aspects of the fusion energy power plant
system.  For MIF systems, the final beam pulse duration is longer,
($>\!\!\unit[1]{\mu s}$), while the desired ion energy is typically lower
($<\!\!\unit[1]{MeV}$).  Thus, the required current could be higher
depending on the drive pulse duration which can vary widely from the
nanoseconds timescale to hundreds of microseconds.\cite{Wurden2016}

Most ion accelerators have average acceleration gradients of
\unitfrac[1-5]{MV}{m} over their length.  Comprising focusing magnets
and acceleration cells (radio-frequency or induction accelerators), they have
transverse dimensions approximately \unit[0.1-0.5]{m}, not including
power supplies, vacuum pumps, and other ancillary equipment.  For
non-fusion applications, the beam requirements tend to be less
demanding. Nevertheless, the state-of-the-art accelerator architectures in these
applications have similar footprint and cost.

This work is motivated by the use of microfabrication based approaches
to both miniaturize and reduce system cost. Applying fabrication
approaches widely used to fabricate Microelectromechanical systems
(MEMS), waferscale fabrication has the potential of reducing cost and
increasing performance.\cite{Shi2011} In particular, fabrication of
small orifices in the millimeter range with submicron lithography
enables high precision and high electric fields at modest voltages. In
order to generate high power beams, we propose to accelerate many
parallel beams in a common accelerator structure and vacuum
system. Utilizing MEMS technology, the beamlets can be densely packed,
leading to a lower-cost and reliable accelerator architecture for the
applications above.

The idea of a scaled down high energy physics accelerator architecture
was presented by Maschke,\cite{Maschke_1979} motivated by the need to
generate high current and brightness beams for ion-driven IFE. He
proposed a Multiple Electrostatic Quadrupole Array Linear Accelerator
(MEQALAC) structure consisting of electrostatic quadrupoles (ESQs)
and radio-frequency (RF) acceleration. The current limit for an ESQ
focusing lattice is given by
\begin{equation}
 i_{max} = 4\times 10^{-12} a E v
\end{equation}
where $a$ is the beam aperture (distance to the focusing electrode
pole tip), $E$ the maximum electric field, and $v$ the ion beam
velocity.\cite{Hogan, Faltens} The peak electric field is bound by
the breakdown field of the quadrupole structure. Measured breakdown
fields stay constant or increase when scaling down the gap distance.\cite{Slade, Faltens} The current density averaged over the focusing
structure increases inversely with $a$ and the total current can be
increased by accelerating and focusing multiple closely-packed
beamlets in separate focusing channels. Although magnetic quadrupoles
offer stronger focusing than ESQs at higher beam
velocities, they do not scale well to smaller sizes, i.e., the current
density needed to create the magnetic field increases when scaling
down the quadrupole size. In practice, the smallest useful size for the
quadrupole structures will be limited by alignment and fabrication
errors of the structure. MEMS technology pushes the boundaries of
these fabrication errors into the sub-micrometer range and, therefore,
allows drastic improvements in transportable currents. In fact, the
theoretical current density transport limits achievable with an
aperture of the order of $\unit[100]{\mu m}$, a unit cell size of
$\unit[500]{\mu m}$, and an applied field of
$E = \unitfrac[10^8]{V}{m}$ (taken as 50\% of the breakdown limit from
Fig.~4 in Ref.~\citenum{Slade}) will be $\sim\unitfrac[2]{A}{cm^2}$ for Xe
ions and $\unitfrac[22]{A}{cm^2}$ for H ions, both at
\unit[10]{keV}. Normal ion sources produce of the order of
$\unitfrac[10]{mA}{cm^2}$ beam currents for Xe. For hydrogen,
$\unitfrac[0.6]{A}{cm^2}$ has been achieved.\cite{Skalyga} The ion
source will therefore be the limiting factor in most applications, and
bunching, funneling, or other methods of increasing the current density
might be utilized to overcome this limitation.

\section{Concept}

\begin{figure}[ht!]
  \centering
  \subfigure[RF stack]{  \label{fig:rf-concept}
    \includegraphics[width=0.35\linewidth]{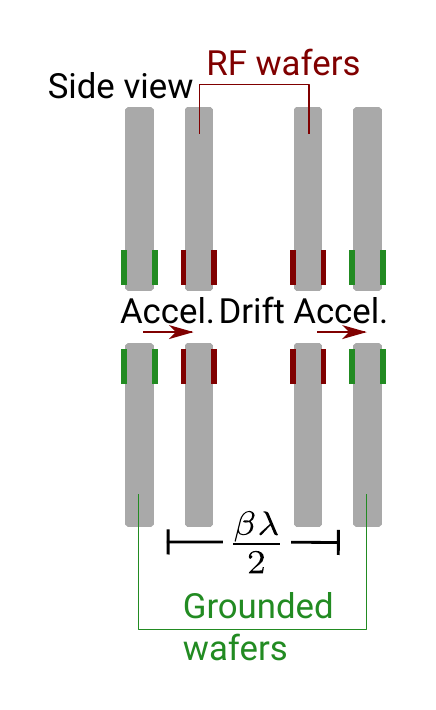}}
  \hfill
  \subfigure[ESQ]{  \label{fig:esq-concept}
\includegraphics[width=0.55\linewidth]{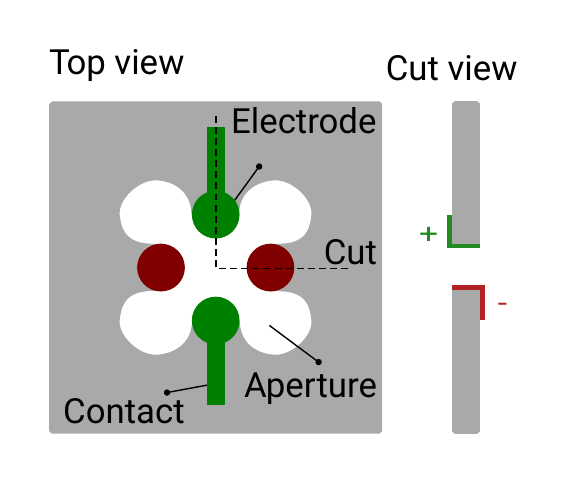}}
\caption{(a) RF-acceleration concept using four RF
  wafers shown for a single beamlet. The outside wafers are grounded
  and RF voltage is applied to the inner wafers, creating
  acceleration fields and a drift region. Particles can be accelerated
  twice, if the drift region length is chosen to match the ion
  velocity $\beta$ and half the wavelength, $\lambda$, of the RF. (b) ESQ concept implemented in a
  single wafer for a single beamlet. The quadrupole length is defined
  by the side-wall coating as shown in the cut view.}
\end{figure}

The proposed accelerator structure consists of two main components: RF
units and ESQ doublets. The RF units will provide the acceleration for
the ion beam bunches, and the ESQs will provide transverse focusing
along the accelerator structure. Each RF unit consists of two vacuum
gaps between ring electrodes that define acceleration and field-free
drift regions. The field-free drift gap is set so that
\begin{equation}
 d = \frac{\beta \lambda}{2},
\end{equation}
where $\lambda$ is the RF wavelength and $\beta =\frac{v}{c}$, and $d$
is the center-to-center distance between adjacent acceleration
gaps. Successive gaps can easily take into account the increasing
particle $\beta$ due to acceleration. Thus particles see nearly
identical acceleration fields in adjacent gaps since the RF phase
advances by $\pi$ (see Fig.~\ref{fig:rf-concept}). For each RF unit,
the ion bunches enter and leave the unit at ground potential. Components
consist of either silicon wafers or printed circuit boards (PCB). For
this paper, we will refer to both as wafers independent of the chosen
implementation. All RF wafers share the same design making batch
fabrication possible: each beamlet corresponds to a through hole
aperture in the wafer that has a ring electrode at its entrance and
exit on the surface of the wafer. In silicon, this would be a deposited
metal ring, and in our PCB implementation, the on-board copper is
utilized for this purpose. To create an RF unit, four wafers are
stacked together. Both sides of the outer wafers are grounded and the
four sides of the inner wafers are connected to a common RF source
(see Fig.~\ref{fig:rf-concept}). We have chosen acceleration gaps of
\unit[1.4]{mm} for the experiments reported here. Precision washers
are used between the wafers to define the gap distances.

For transverse focusing, each ESQ consists of two pairs of electrodes
which are biased to $\pm V_Q$. To implement ESQ components, we form
four electrodes around each beam aperture with electrical
connections to the front and back of each wafer for the positive and
negative voltages (see Fig.~\ref{fig:esq-concept}). The ESQ structure
is therefore completely contained in a single wafer. A single ESQ
wafer focuses the beam in one direction and defocuses the beam in the
other. We therefore use two ESQ wafers to form a focusing doublet to
provide an overall focusing effect on the beam.

\section{Fabrication}

For initial proof-of-concept experiments, we utilize FR-4 PCB to
fabricate the RF and ESQ wafers. We are also exploring the fabrication
of quadrupole structures using 3D-printing techniques as well as
quadrupoles based on silicon wafers.

\begin{figure}[ht!]
  \centering
  \includegraphics[width=\linewidth]{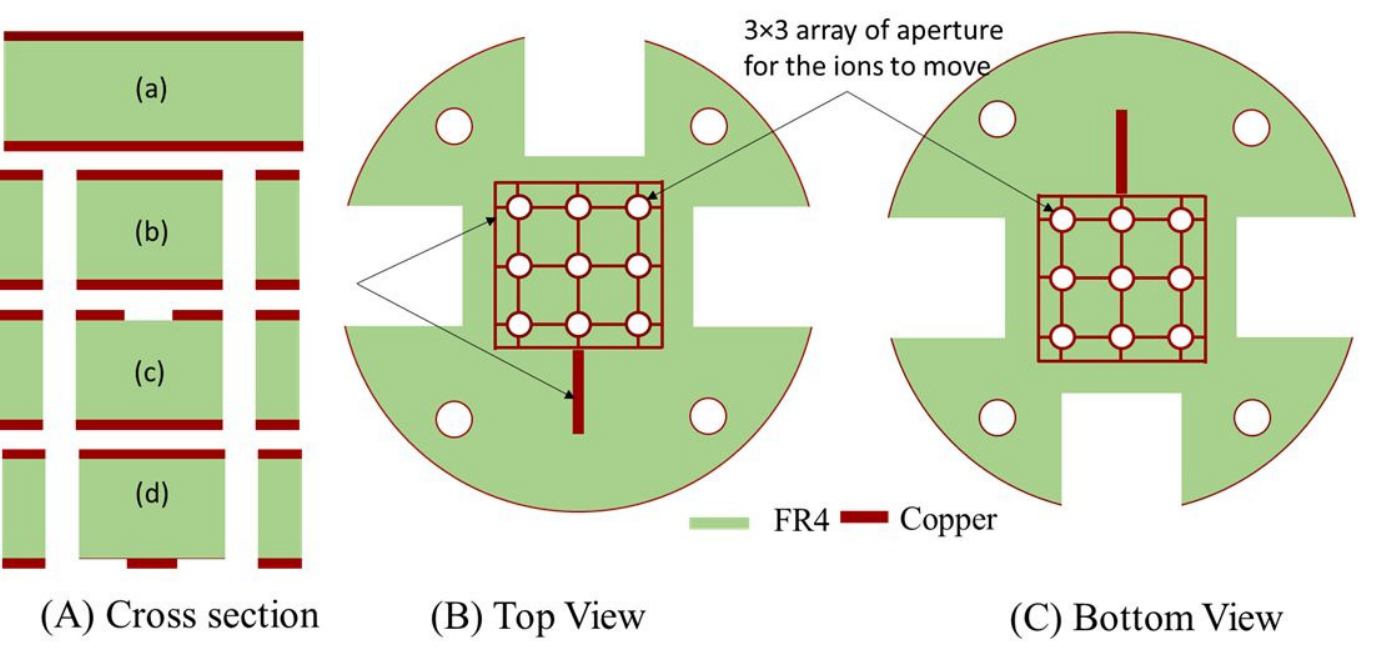}
  \caption{(A) PCB fabrication procedure for RF
    wafers: (a) The process starts with a double
    clad, 0.028 inch, 1 oz. FR-4 board that is cut in the shape of a 4 inch
    wafer. (b) Holes are cut into the PCB. (c) The top metal layer is
    patterned using laser after alignment with fiducials. (d) The bottom
    metal is patterned using a laser after alignment with fiducials. (B) Top view of
    the fabricated RF wafer. (C) Bottom view of the fabricated RF
    wafer.}
  \label{fig:rf-fab}
\end{figure}

Using laser micromachining (LPKF ProtoLaser U4), top and bottom metal
layers are patterned and holes are drilled through the
PCB. Alignment between top and bottom is achieved by using an
integrated vision system and pre-fabricated alignment fiducials. Steps
of the process to fabricate RF wafers are given in
Fig.~\ref{fig:rf-fab}. In this process, we start with a FR-4 based board
that has copper on both sides as seen in the cross section
[Fig.~\ref{fig:rf-fab} (A)]. The circular holes are created using a
laser tool. Then laser cutting is used to define top and bottom metal
routing. The top and bottom views of the fabricated RF wafer are also
shown in Fig.~\ref{fig:rf-fab}.

\begin{figure}[ht!]
  \centering
  \includegraphics[width=0.5\linewidth]{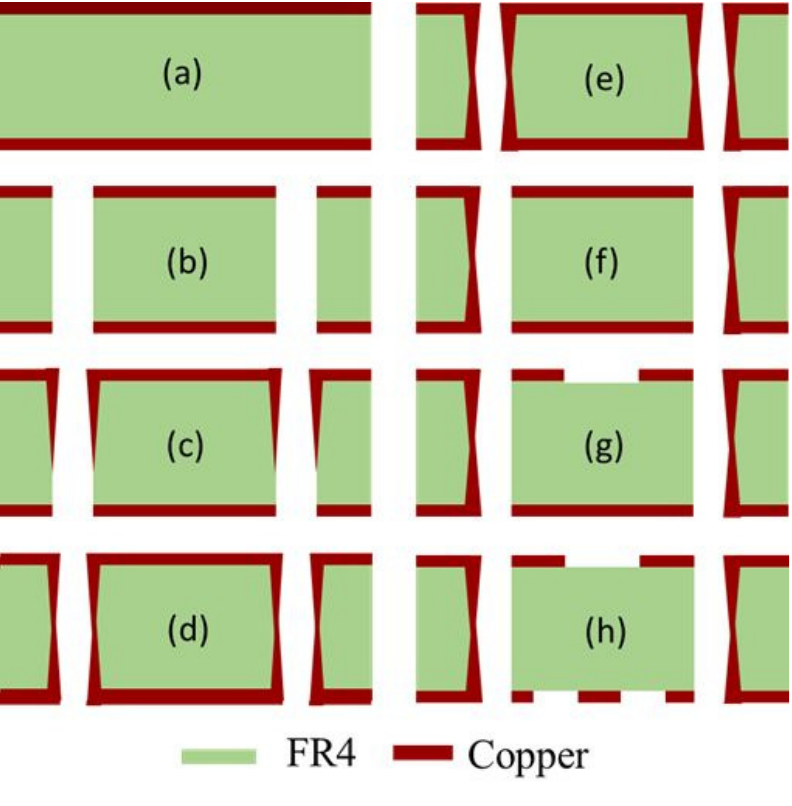}
  \caption{PCB fabrication procedure for ESQ wafers:
    (a) The process starts with a double clad, 0.028”, 1 oz FR-4 board
    that is cut in the shape of a 4 inch wafer. (b) Holes are cut into
    the PCB. (c) \unit[500]{nm} thick Cu is deposited by conformal
    evaporation from top of the wafer. (d) \unit[500]{nm} Cu is
    evaporated from bottom of the wafer. (e) Metal is electroplated
    for better coverage of the sidewalls (f) Isolation cut with a
    laser to remove part of the sidewall over which no metal is
    desired. (g) Top metal layer is patterned using laser after
    alignment with fiducials. (h) Bottom metal is patterned after
    alignment with fiducials. }
  \label{fig:esq-fab}
\end{figure}

The main steps of the process to fabricate ESQ wafers are given in
Fig.~\ref{fig:esq-fab}.  The FR-4 based board has copper on both sides
and holes are then created using the laser tool. As the holes in the
PCBs are created using a scanned laser beam rather than a milling
tool, arbitrary hole shapes can easily be realized. After cutting the
holes, a copper layer ($\unit[1]{\mu m}$ thick) is evaporated onto the
board in a conformal evaporator with a rotating chuck system on both
sides. For better sidewall coverage, copper is also electroplated on
top of the evaporated copper, which serves as a seed layer for the
electroplated copper.  The laser is then used to remove the metal on
certain parts of the sidewalls of the holes to isolate the
electrode. Finally, the laser is used again to define top and bottom
metal routing.

\section{Experimental Setup}

\begin{figure}[ht!]
  \centering
  \includegraphics[width=\linewidth]{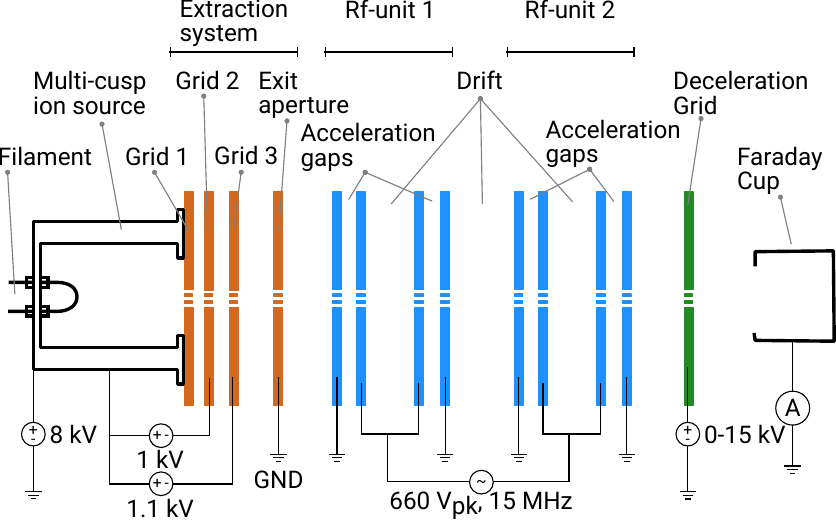}
  \caption{The experimental setup consists of a
    filament driven ion source with a three grid extraction system, that provides a
    $3\times3$ array of beamlets. The source is biased at high voltage
    for beam extraction. A RF unit cell (shown) or an ESQ doublet is
    inserted, followed by a Faraday-cup to measure the
    beam current. Alternatively a scintillator and a CCD camera are
    used to image the beam.}
  \label{fig:setup}
\end{figure}

To test the major components of this technology, a compact test stand
has been set up. A filament driven, multi-cusp ion source \cite{Ji16}
is used to generate plasma. The source operates using argon or
helium at a pressure in the plasma chamber of around
\unit[6]{mTorr}. To create the plasma, we rely on continuous gas flow
rather than periodic gas puffs. The filament is then operated at
several amperes and a \unit[-100]{V} arc pulse is applied for
$\unit[300]{\mu s}$ to ignite the plasma. Ions are extracted during
the arc pulse by floating the source body to a high voltage
($\unit[\leq 15]{kV}$). Along with the plasma density, a
three-electrode extraction system determines the plasma meniscus
shape, extracted current density, and beamlet envelope trajectory at
the high voltage acceleration gap (see Fig.~\ref{fig:setup}). The
plasma facing electrode (grid 1 in Fig.~\ref{fig:setup}) is not
electrically connected to a fixed voltage source and, therefore, floats
to the plasma potential during operation. A voltage of $\unit[-50]{V}$
has been measured during plasma operation. The second electrode (grid 2) is
used to extract the ion beam and is biased at a negative voltage
relative to the source body.  A third electrode (grid 3) has been implemented
that can be utilized to gate the extracted beam and short, uniform ion
pulses of $\unit[4]{\mu s}$ bunch duration with no electron pre-pulse
have been achieved. However, in the work reported here, we do not
utilize the gating mechanism and the third electrode is biased
slightly more negative than the second grid, so that ions are
extracted continuously during the arc-pulse. The ions then gain their
remaining kinetic energy when they are accelerated from the source
potential to a grounded exit aperture.

\begin{figure}[ht!]
  \centering
  \includegraphics[width=\linewidth]{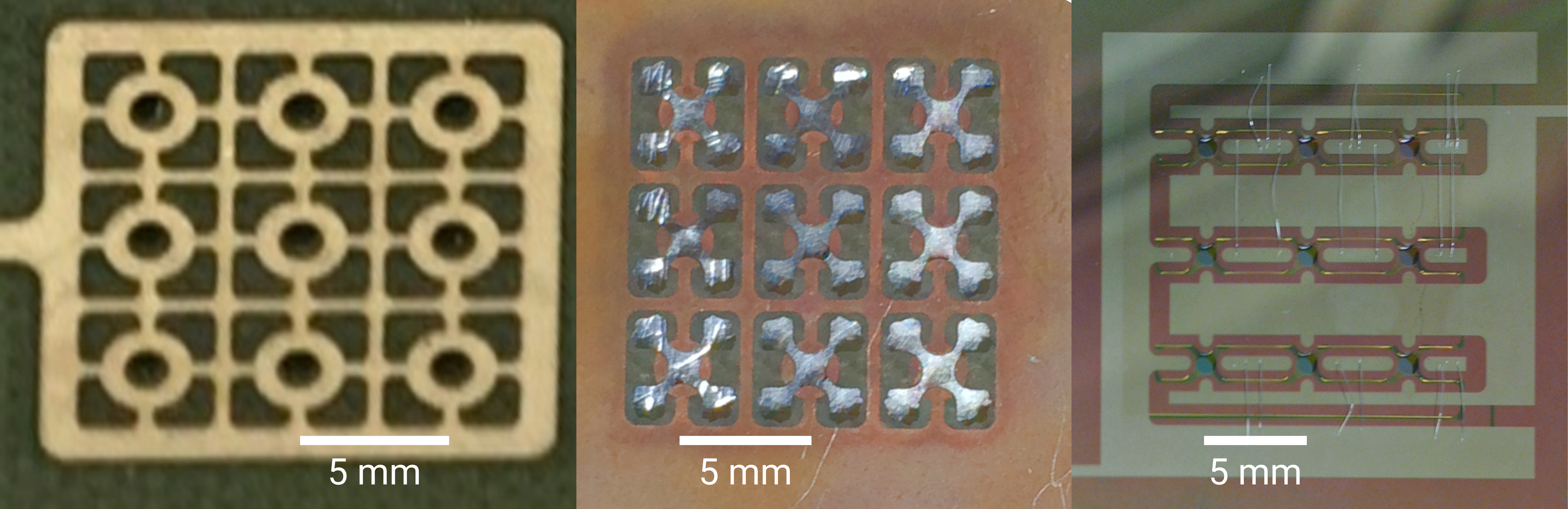}
  \caption{Different implementations of a
    3$\times$3 array with a \unit[5]{mm} spacing, matching our
    extraction system at the source. Left to right: PCB RF wafer, PCB
    ESQ wafer, silicon ESQ wafer. Results reported in this paper used
    the PCB implementation shown here.}
  \label{fig:implementation}
\end{figure}

To demonstrate multiple parallel beamlets, we use a multi-aperture
extraction system. In our prototype, a $3\times3$ array of beam
extraction apertures is implemented for each electrode, where each
extraction aperture has a \unit[0.5]{mm} diameter and the electrodes are
positioned with an inter-hole pitch of \unit[5]{mm}, see Fig.~\ref{fig:implementation}.

After the ions leave the beam extraction section, we utilize the beam
to test our accelerator components. For the RF experiment, the test
structure consisted of two RF units. The drift gap length was designed
to match an \unit[8]{keV} argon ion beam, a driving RF frequency of
\unit[15]{MHz}, and an acceleration voltage of \unit[1]{keV} per gap. We
summarize the implemented distances in our two RF-unit (four
acceleration gaps) setup in Table~\ref{tab:drifts}.
\begin{table}
\caption{\label{tab:drifts}Drift distances used between RF
  acceleration gaps designed for \unit[15]{MHz}, \unit[1]{kV} RF signal, and
  a \unit[8]{keV} ion beam. A RF unit consists of two acceleration gaps.}
\begin{ruledtabular}
\begin{tabular}{ll}
Drift gap number & Drift length (\unit{mm}) \\
1 &4.00 \\
2 &4.37\\
3 &4.73\\
\end{tabular}
\end{ruledtabular}
\end{table}

The RF high-voltage for our experiments is generated by amplifying the
output of a signal generator (Agilent 33520A) to an output power of at
most \unit[1]{W} (Mini Circuit ZHL-2-8) where it can be used to
directly drive the gate of an RF metal–oxide–semiconductor
field-effect transistor (MOSFET) and excite the tuned resonant circuit
at the same frequency resulting in another gain of $\approx 40$.  This
design is based on the RF generator for the buncher for the 88-inch
cyclotron at Lawrence Berkeley National Laboratory (LBNL).\cite{Todd-cyclotron2013} The losses in the MOSFET and the RF wafer
capacitance ($\unit[\sim\!\!1]{pF}$ per gap) are the dominant loads. The
frequency range used for the experiments presented here is
\unit[10-20]{MHz}. A peak amplitude on the PCB of up to \unit[660]{V}
has been measured.

In the absence of a pre-bunching (or chopper) section, the ion source
will deliver a constant ion current over many RF periods. Therefore,
some ions are expected to be accelerated and others decelerated. A
retarding potential analyzer is used to measure the beam energy
distribution. This is implemented by adding a biased grid after the
RF units followed by a Faraday cup to measure the beam current. By
scanning the grid voltage, the Faraday-cup at the end of the beamline
selectively detects the current of ions with a kinetic energy higher
than the bias voltage on the grid. For a larger accelerator structure,
only a pulsed beam can be transported through the accelerator
structure, since the focusing elements as well as the drift gaps will
be tuned for a single ion energy. Without bunching the beam, one can
expect a transport limit of about 10\% of a continuous beam to be
accelerated through an RF accelerator. This transport limit is given
by the requirement for longitudinal phase stability.\cite{Wangler2008}

The ESQ wafers are tested in a similar fashion. Instead of the
RF units, a single ESQ wafer or a doublet is mounted downstream of the
ion source. We then use a scintillator (RP 400 plastic scintillator)
and a fast image intensifying camera (Princeton Instruments) to
observe the beam transverse distribution. Since a filament driven ion
source is being used, light from the filament is also detected by the
camera. By looking at the scintillator at an angle, we avoid
overlapping the light from the filament with light output from the
ion beam hitting the scintillator. As both have roughly the same
amplitude, this avoids the need for background subtraction. Voltage
scans on the ESQ electrodes then result in beam deflection that can be
measured.

\section{Results}

\begin{figure}[ht!]
  \centering
  \includegraphics[width=\linewidth]{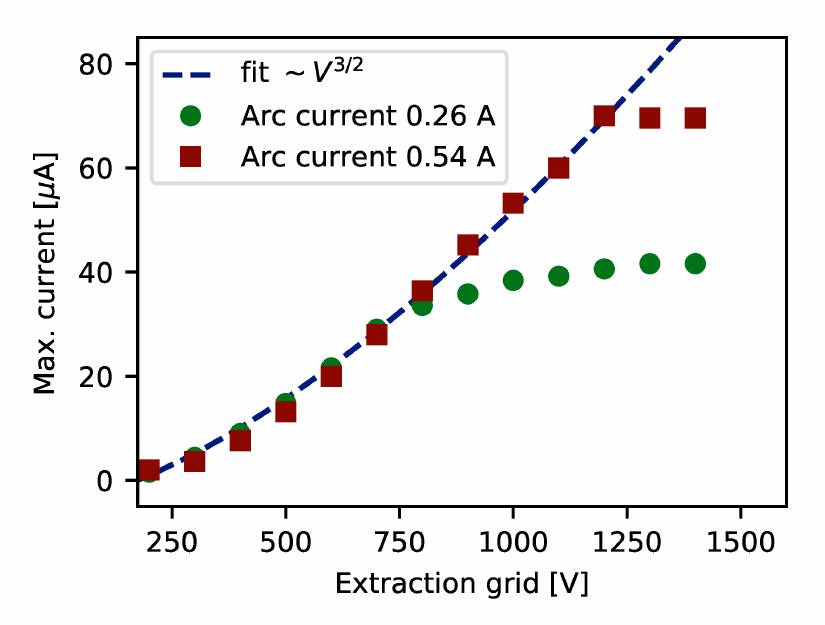}
  \caption{Scanning the extraction grid voltage the
    source shows space-charge limited extraction behavior before the
    emission saturates depending on the arc current.}
  \label{fig:source}
\end{figure}

To characterize the ion source, we scanned the extraction voltages of
the ion source. The source was operated at \unit[6]{mTorr} using
argon. The filament was on for \unit[7]{s} to create a sufficiently
high, stable filament temperature. Then the arc voltage was pulsed to
\unit[-100]{V} for \unit[0.3]{ms}. The source was floated at
\unit[8]{kV} and then extraction voltages on the second and third
electrode were scanned. Here, the third electrode was always held
\unit[100]{V} more negative than the second
electrode. Figure~\ref{fig:source} shows the resulting beam current
measured from the source without any RF or ESQ units present. The
ion current increases as $V^{\frac{3}{2}}$ as expected from the
Child-Langmuir law of space-charge limited extraction and then at
higher extraction voltage, depending on the plasma density, changes to
an emission limited regime. Operating the source in the emission
limited regime will result in more shot-to-shot variation, since the
output level will depend on the gas pressure in the source chamber,
filament conditions, etc. Therefore, an operating point at an
extraction voltage of \unit[1000]{V} was chosen for an arc current of
$\unit[0.6]{A}$, which provided very stable source performance. The
plasma facing electrode has $\unit[0.5]{mm}$ diameter holes, thus the
local extracted current density is $\unitfrac[30]{mA} {cm^2}$ at the
operating point.

\begin{figure}[ht!]
  \centering
  \includegraphics[width=\linewidth]{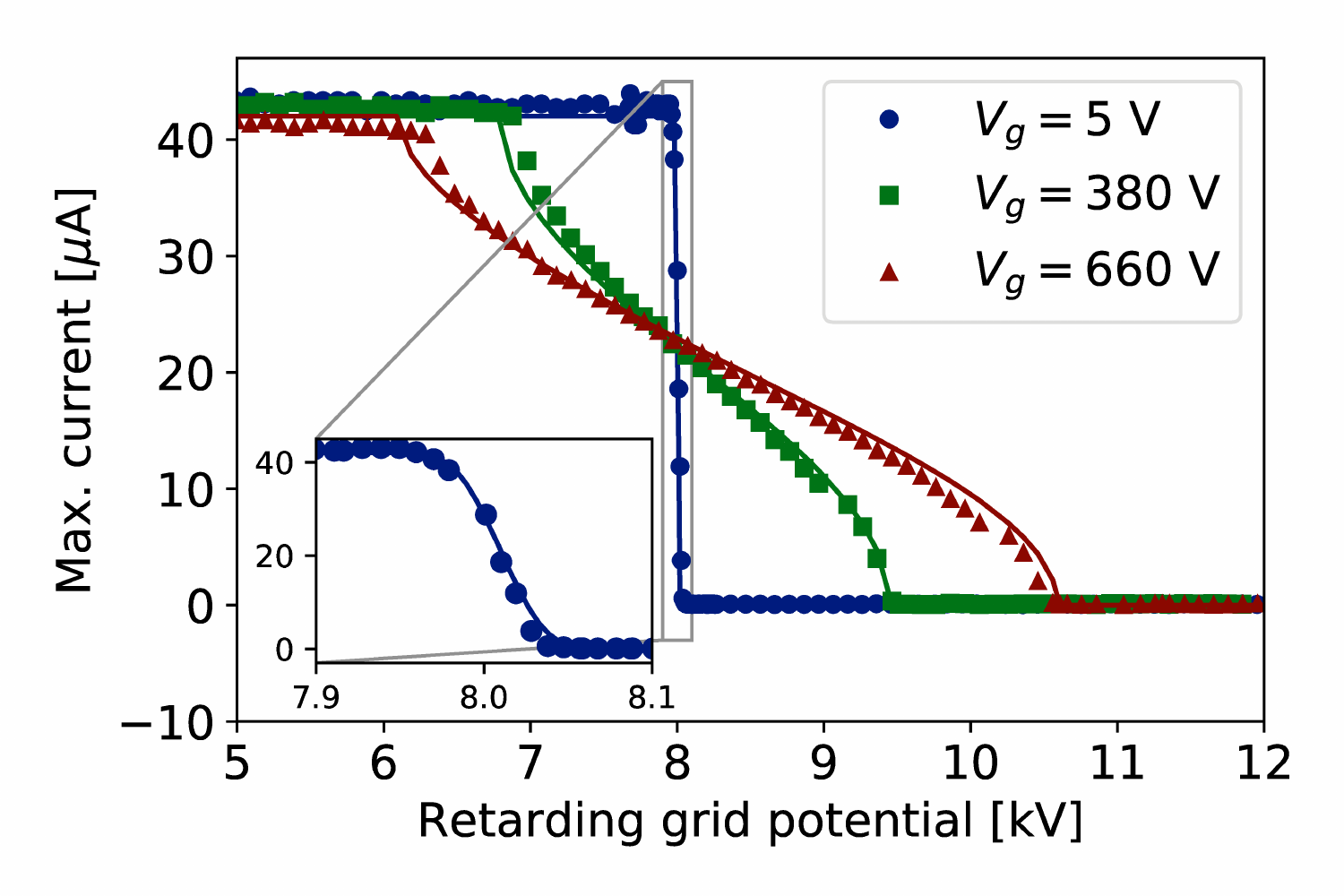}
  \caption{Result from two RF units operating at
    different RF amplitudes are shown. The RF frequency was
    \unit[15.03]{MHz}. The solid lines show 1-D simulation results. The inset shows
    a zoom-in of a run with almost no RF applied, characterizing the beam
  energy. The energy spread in the inset is due to use of a mesh for
  the deceleration stage.}
  \label{fig:rf-result}
\end{figure}

Next, two RF units were tested. The beam energy profile was
measured for three different conditions: (1) minimum RF amplitude
$\approx \unit[5]{V}$ (lowest setting on the RF generator) (2)
RF amplitude at \unit[380]{V} amplitude and (3) RF amplitude at
\unit[660]{V} amplitude. The results are shown in
Fig.~\ref{fig:rf-result} together with simulated results (solid
lines). The simulation uses a 1-D model to calculate the beam energy of
macro particles with nanosecond separation along the beam pulse. These particles are
traced to the center of the acceleration gap where the particle energy
is changed according to the RF phase at that time. Once the particles
reach the position of the Faraday cup, a perfect energy filter is
assumed and the charge of particles higher than the simulated
retarding-grid voltage is integrated. The simulated
results are based on the applied RF frequency and amplitude, the ion energy,
mass and current, as well as the acceleration gap positions. The
results of the simulation are then scaled to the initial beam
current (the only free parameter in the simulation code). Experimental and simulation
results using several RF gaps and the simulation program itself will
be discussed in detail in a future publication.

At the lowest RF settings of \unit[5]{V}, an energy spread of about
\unit[15]{eV} is visible (see the inset of
Fig.~\ref{fig:rf-result}). This spread can be attributed to an
intrinsic broadening of our diagnostic and fits well with a model from
Sakai and Katsumata\cite{Sakai1985} that predicts a \unit[15]{eV}
energy resolution for a rectangular mesh of 90 lines/inch and
$\unit[5.5 \times 10^{-3}]{inch}$ as used in our experiment.  The ion
energy spread of a filament driven ion source is small,\cite{Lee1996} 
on the order of \unit[5]{eV}. Since the energy spread within an
RF-bucket will be more than an order of magnitude larger, the energy
spread at the source is not explicitly included in the model.  The
measured injected beam energy is \unit[8.01]{keV} based on the
retarding potential scan. For the other cases, we measured a wider
energy spread. As can be seen, some particles arrive at the correct
time to achieve full acceleration in each gap, whereas others arrive
at different phases of the RF and are either significantly
accelerated, experience little or no acceleration or are significantly
decelerated. This is to be expected, since we do not inject a pulsed
beam but use a long pulse in regards to the RF frequency.  In a linear
accelerator, the stable longitudinal bucket occupies about 10\% of the
RF wave.\cite{Wangler2008} In the absent of bunching, as in our
present setup, we expect this fraction of the DC injected beam to be
captured and accelerated through many gaps. This is in agreement with
the data shown in Fig.~\ref{fig:rf-result}. The measured energy
distribution fits very well with our 1-D model that simulates a
continuous ion beam.  As one can see, the maximum beam energy is
roughly given by the injected beam energy plus four times the maximum
RF amplitude. The fact that the measured energy is slightly lower can
be attributed to the fact that the gap distances were designed for a
slightly higher RF amplitude.  For a RF accelerator consisting of many
stages, one would inject a pulsed beam or use a pre-buncher module to
only inject particles at the right phase of the RF.

\begin{figure}[ht!]
  \centering
  \includegraphics[width=\linewidth]{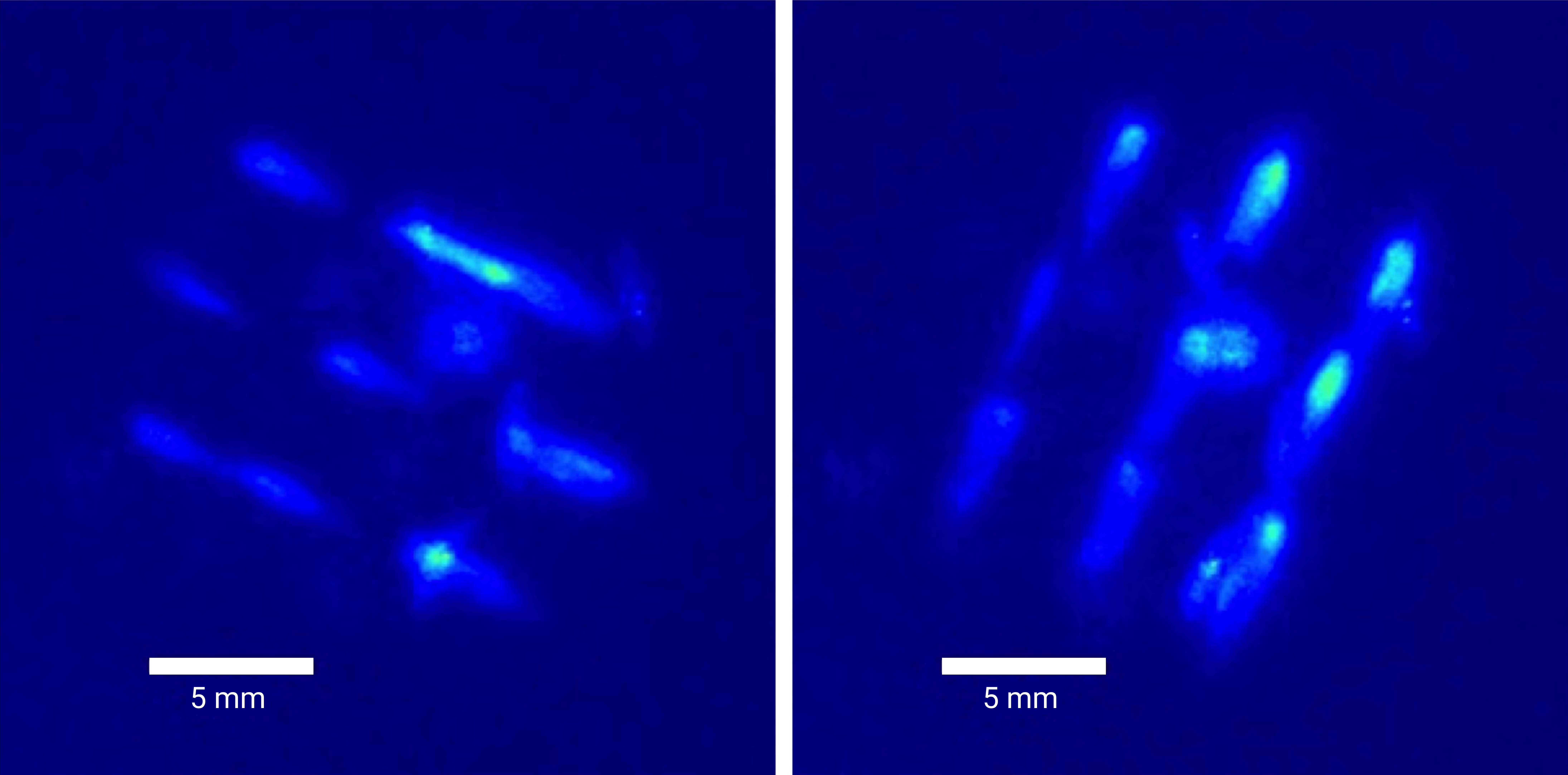}
  \caption{ESQ focusing effect for positive and
    negative bias at \unit[280]{V} recorded using a scintillator and a
    CCD camera. As expected focusing in one direction and defocusing
    in the other direction is observed. Common background light
    from the filament of the source is visible in both images.}
  \label{fig:esq-result}
\end{figure}

\begin{figure}[ht!]
  \centering
  \includegraphics[width=0.8\linewidth]{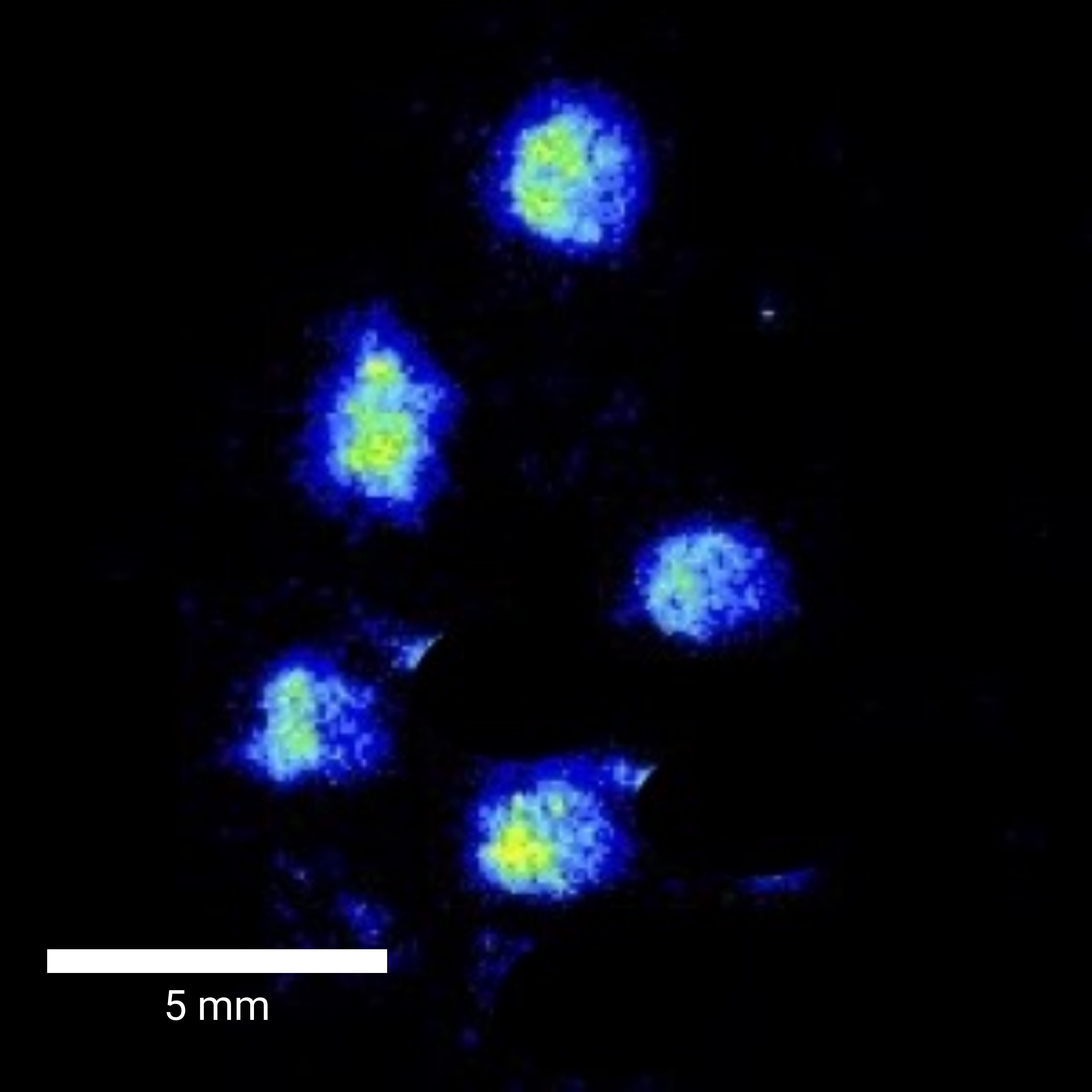}
  \caption{A focusing effect for two ESQs using
    alternating gradients at \unit[150]{V} is shown.  To reduce the background
    light from the ion source entering the CCD camera optics, four
    beams in the 3x3 array were masked.}
  \label{fig:esq-result-AG}
\end{figure}

For the ESQ tests, we utilized a single ESQ wafer and demonstrated the
characteristic elliptical deformation of a round beam that is the result of
focusing the beam in one plane and at the same time defocusing the
beam in the orthogonal plane (see Fig.~\ref{fig:esq-result}).
Combining two ESQs into a doublet then allows the beam to be focused
in both directions. To demonstrate this, we chose voltages of an ESQ
doublet such that the initial round beam is again focused to a round
beam after passing two ESQs, as shown in
Fig.~\ref{fig:esq-result-AG}. The radius of the focused round beam can
be influenced by the applied voltages as expected.

\section{Discussion}

In this paper, we have demonstrated the functioning of basic components
needed to implement a millimeter-scale MEQALAC formed from FR-4
PCB. The tested structures were made using a laser
microfabrication tool that enables rapid prototyping for testing
devices but is not the best in fabrication precision. The laser cutter
provides resolution of $\unit[10-20]{\mu m}$ over orifices of 100s of
microns. Furthermore, the accuracy over a 4-inch PCB is made
worse by the travel inaccuracies of the stages used to move the laser
beam and the substrate.  Another drawback of the laser
microfabrication approach is that the speed of manufacturing is
limited owing to the beam-based fabrication, much like exposure times
in electron-beam lithography. However, our accelerator component
design may be adapted to optical lithography fabrication in
combination with silicon micromachining, which can lead to
$\unit[0.5]{\mu m}$ accuracy over a 4-inch wafer, while being
fabricated in wafer-batches. These approaches are currently under
development in our labs. By scaling our devices to be fabricated using
silicon micromachining approach, we can achieve much greater uniform
focusing elements leading to higher overall accelerator current
densities.

The results reported here open a path to scaling to much higher ion
beam currents and beam energies, which has the potential for ion beam
systems in a broad series of established application areas where
system cost and foot print can be reduced drastically. These results
could also open up new application areas where system costs today make
ion beam applications prohibitively expensive. Driver beams
requirements to heat plasmas or targets for fusion energy are among
the most demanding. Massively scaled MEMS multi-beam RF accelerators
might offer a promising path for low cost ion beam fusion drivers.
Our experiments have focused on accelerator technology and the
demonstration of a scalable architecture to a large number of beams at
a low cost.  We have injected and transported
$\unitfrac[5-10] {\mu A}{beam}$ in a $3\times3$ array.  The pitch of
the array is \unit[5]{mm}, thus, the average current density over a
$\unit[5\times5]{mm}$ unit cell is: $\unit[5-10]{\mu A}$ /
$\unit[0.25]{cm^2} = \unitfrac[20-40]{\mu A}{cm^2}$.  The quadrupole
focusing results and the acceleration of the beam through two RF-units
(two gaps per unit) show encouraging results, as shown in the
preceding section.

The achieved beam current averaged over the array unit cell is still a
long way from what is required for a fusion driver.  To gauge the
remaining challenges, consider a xenon driver beam (singly ionized) at
\unit[1]{MeV} for an unspecified fusion target requiring \unit[1]{MJ}
of beam energy in a $\unit[1]{\mu s}$ pulse.  We consider half of the
surface area, $2\pi$ steradians, of a \unit[5]{m} radius fusion chamber
will be occupied by about 2000 MEMS accelerators comprised of \unit[30]{cm}
diameter wafers within a vacuum enclosure.  Each accelerator module
must deliver \unit[500]{A} of ions, and the peak current density
averaged over a beamlet unit cell must be ca
$\unitfrac[1.6]{A}{cm^2}$.  To estimate the achievable beam parameters
in a MEMS accelerator using demonstrated ion source technology, we
assume that the ion sources deliver $\unitfrac[100]{mA}{cm^2}$, as
achieved previously for argon beams.\cite{Kwan2004} Assuming an
aperture radius of \unit[0.5]{mm} (as for the PCB ESQs tested
above) and quadrupole electrode voltage of \unit[2.5]{kV}, the peak
field between the electrodes is \unitfrac[100]{kV}{cm}, a conservative
design specification.  Thus, the current density over a beamlet unit
cell would be ~$\unitfrac[20]{mA}{cm^2}$.  This is lower by a factor
of about $80$ than the needed current density for a driver. (For
fusion target designs accommodating higher ion kinetic energy, the
required current density decreases proportionally.) Strategies for
reducing this could be a combination of source and injector
improvements,\cite{Grote-injector2003} improved beam capture
efficiency into the RF structure, and macro-bunch compression after
the RF accelerator.\cite{Qin-compression2005}  These will be explored
in future work and we note that they have been explored already for
heavy ion driven inertial fusion system designs.

We envision this technology to be applicable for ion beams in the
\unit[100]{keV} to several \unit{MeV} range, with average beam current
densities up to $\unitfrac[1]{mA}{cm^2}$. The technology can also be
used for lower energies. However, below \unit[100]{keV} it will be
more effective to use a single high voltage gap to accelerate the ions
directly. Above \unit[100]{keV} beam energy, the advantages of having
inexpensive components and lower peak voltages will be more and more
important. For this technology to be competitive, we believe that an
RF amplitude of several kilovolts is needed per acceleration gap. This
way, we will be able to achieve gradients of \unitfrac[1]{MV}{m} for
the accelerator structure at frequencies in the \unit[50]{MHz}
range. To accomplish these high gradients, we are currently
investigating the use of onboard resonators with a high Q that
already have been shown to produce the required voltages. First
prototypes are currently being designed and we will test these devices
in the coming months to be able to integrate most of the RF stack onto
the wafers. Switching from PCB to silicon will also provide a path for
mass fabrication and better manufacturing precision.

In comparison to other current technologies, for example radio
frequency quadrupoles (RFQs), MEQALACs can achieve higher total
transported currents utilizing multiple beamlets. The overall energy
efficiency of a MEQALAC also scales favorable in comparison to RFQs. A
detailed comparison of these two technologies can be found in the
paper of Urbanus \textit{et al.}\cite{Urbanus1990} Another
interesting metric is the effective transportable current density in
respect to the cross section of the device (instead of the beam
diameter). For state of the art RFQs of the order of
$\unitfrac[70]{\mu A}{cm^2}$ (\unit[50]{mA} in a $\unit[30]{cm}$
diameter vacuum chamber) has been established.\cite{Ratti-SNS2002} Our
prototype MEQALAC design delivers around $\unitfrac[40]{\mu A}{cm^2}$
($9\times\unit[10]{\mu A}$,
$\unit[1.5]{cm}\times\unit[1.5]{cm}$). However, denser packaging can
increase these by a factor of 10-30 in PCB implementations and the
use of MEMS and silicon substrates will allow scaling to even higher
current densities of the order of $\unitfrac[1]{mA}{cm^2}$.

\section{Conclusion and Outlook}

We have shown that a compact RF unit without a resonant cavity can be
used to accelerate an ion beam. For this proof of concept, PCB
structures have been used, whereas for a final accelerator we envision
the use of silicon micromachined wafers. This will allow smaller
beamlets packed to a higher density on a wafer for increased effective
beam-current densities. Furthermore, focusing elements in the form of
ESQs will be added to the accelerator to allow for beam transport and
refocusing of the beam along the beamline. Once we achieved
integration of RF units and ESQ doublets, we will start scaling the
concept to a stack of 10 RF units. This will allow us to examine
scaling laws to higher energies. Initial simulations indicate that
higher energies and currents will be achievable.

\begin{acknowledgments}
We are grateful for insightful discussions with Andris Faltens (LBNL).

This work was supported by the Office of Science of the US Department
of Energy through the ARPA-E ALPHA program under contract
DE-AC02–05CH11231 (LBNL).
\end{acknowledgments}


\begin{thebibliography}{22}%
\makeatletter
\providecommand \@ifxundefined [1]{%
 \@ifx{#1\undefined}
}%
\providecommand \@ifnum [1]{%
 \ifnum #1\expandafter \@firstoftwo
 \else \expandafter \@secondoftwo
 \fi
}%
\providecommand \@ifx [1]{%
 \ifx #1\expandafter \@firstoftwo
 \else \expandafter \@secondoftwo
 \fi
}%
\providecommand \natexlab [1]{#1}%
\providecommand \enquote  [1]{``#1''}%
\providecommand \bibnamefont  [1]{#1}%
\providecommand \bibfnamefont [1]{#1}%
\providecommand \citenamefont [1]{#1}%
\providecommand \href@noop [0]{\@secondoftwo}%
\providecommand \href [0]{\begingroup \@sanitize@url \@href}%
\providecommand \@href[1]{\@@startlink{#1}\@@href}%
\providecommand \@@href[1]{\endgroup#1\@@endlink}%
\providecommand \@sanitize@url [0]{\catcode `\\12\catcode `\$12\catcode
  `\&12\catcode `\#12\catcode `\^12\catcode `\_12\catcode `\%12\relax}%
\providecommand \@@startlink[1]{}%
\providecommand \@@endlink[0]{}%
\providecommand \url  [0]{\begingroup\@sanitize@url \@url }%
\providecommand \@url [1]{\endgroup\@href {#1}{\urlprefix }}%
\providecommand \urlprefix  [0]{URL }%
\providecommand \Eprint [0]{\href }%
\providecommand \doibase [0]{http://dx.doi.org/}%
\providecommand \selectlanguage [0]{\@gobble}%
\providecommand \bibinfo  [0]{\@secondoftwo}%
\providecommand \bibfield  [0]{\@secondoftwo}%
\providecommand \translation [1]{[#1]}%
\providecommand \BibitemOpen [0]{}%
\providecommand \bibitemStop [0]{}%
\providecommand \bibitemNoStop [0]{.\EOS\space}%
\providecommand \EOS [0]{\spacefactor3000\relax}%
\providecommand \BibitemShut  [1]{\csname bibitem#1\endcsname}%
\let\auto@bib@innerbib\@empty
\bibitem [{\citenamefont {Takatsu}(2011)}]{Takatsu2011}%
  \BibitemOpen
  \bibfield  {author} {\bibinfo {author} {\bibfnamefont {H.}~\bibnamefont
  {Takatsu}},\ }\bibfield  {title} {\enquote {\bibinfo {title} {{ITER} project
  and fusion technology},}\ }\href {\doibase 10.1088/0029-5515/51/9/094002}
  {\bibfield  {journal} {\bibinfo  {journal} {Nucl. Fusion}\ }\textbf {\bibinfo
  {volume} {51}},\ \bibinfo {pages} {094002} (\bibinfo {year}
  {2011})}\BibitemShut {NoStop}%
\bibitem [{\citenamefont {Hemsworth}\ \emph {et~al.}(2009)\citenamefont
  {Hemsworth}, \citenamefont {Decamps}, \citenamefont {Graceffa}, \citenamefont
  {Schunke}, \citenamefont {Tanaka}, \citenamefont {Dremel}, \citenamefont
  {Tanga}, \citenamefont {{De Esch}}, \citenamefont {Geli}, \citenamefont
  {Milnes}, \citenamefont {Inoue}, \citenamefont {Marcuzzi}, \citenamefont
  {Sonato},\ and\ \citenamefont {Zaccaria}}]{Hemsworth}%
  \BibitemOpen
  \bibfield  {author} {\bibinfo {author} {\bibfnamefont {R.}~\bibnamefont
  {Hemsworth}}, \bibinfo {author} {\bibfnamefont {H.}~\bibnamefont {Decamps}},
  \bibinfo {author} {\bibfnamefont {J.}~\bibnamefont {Graceffa}}, \bibinfo
  {author} {\bibfnamefont {B.}~\bibnamefont {Schunke}}, \bibinfo {author}
  {\bibfnamefont {M.}~\bibnamefont {Tanaka}}, \bibinfo {author} {\bibfnamefont
  {M.}~\bibnamefont {Dremel}}, \bibinfo {author} {\bibfnamefont
  {A.}~\bibnamefont {Tanga}}, \bibinfo {author} {\bibfnamefont {H.~P.~L.}\
  \bibnamefont {{De Esch}}}, \bibinfo {author} {\bibfnamefont {F.}~\bibnamefont
  {Geli}}, \bibinfo {author} {\bibfnamefont {J.}~\bibnamefont {Milnes}},
  \bibinfo {author} {\bibfnamefont {T.}~\bibnamefont {Inoue}}, \bibinfo
  {author} {\bibfnamefont {D.}~\bibnamefont {Marcuzzi}}, \bibinfo {author}
  {\bibfnamefont {P.}~\bibnamefont {Sonato}}, \ and\ \bibinfo {author}
  {\bibfnamefont {P.}~\bibnamefont {Zaccaria}},\ }\bibfield  {title} {\enquote
  {\bibinfo {title} {Status of the {ITER} heating neutral beam system},}\
  }\href {\doibase 10.1088/0029-5515/49/4/045006} {\bibfield  {journal}
  {\bibinfo  {journal} {Nuclear Fusion}\ }\textbf {\bibinfo {volume} {49}},\
  \bibinfo {pages} {045006} (\bibinfo {year} {2009})}\BibitemShut {NoStop}%
\bibitem [{\citenamefont {Nuckolls}\ \emph {et~al.}(1972)\citenamefont
  {Nuckolls}, \citenamefont {Wood}, \citenamefont {Thiessen},\ and\
  \citenamefont {Zimmerman}}]{Nuckolls}%
  \BibitemOpen
  \bibfield  {author} {\bibinfo {author} {\bibfnamefont {J.}~\bibnamefont
  {Nuckolls}}, \bibinfo {author} {\bibfnamefont {L.}~\bibnamefont {Wood}},
  \bibinfo {author} {\bibfnamefont {A.}~\bibnamefont {Thiessen}}, \ and\
  \bibinfo {author} {\bibfnamefont {G.}~\bibnamefont {Zimmerman}},\ }\bibfield
  {title} {\enquote {\bibinfo {title} {Laser compression of matter to
  super-high densities: Thermonuclear ({CTR}) applications},}\ }\href {\doibase
  10.1038/239139a0} {\bibfield  {journal} {\bibinfo  {journal} {Nature}\
  }\textbf {\bibinfo {volume} {239}},\ \bibinfo {pages} {139--142} (\bibinfo
  {year} {1972})}\BibitemShut {NoStop}%
\bibitem [{\citenamefont {Bangerter}, \citenamefont {Faltens},\ and\
  \citenamefont {Seidl}(2013)}]{Bangerter}%
  \BibitemOpen
  \bibfield  {author} {\bibinfo {author} {\bibfnamefont {R.~O.}\ \bibnamefont
  {Bangerter}}, \bibinfo {author} {\bibfnamefont {A.}~\bibnamefont {Faltens}},
  \ and\ \bibinfo {author} {\bibfnamefont {P.~A.}\ \bibnamefont {Seidl}},\
  }\bibfield  {title} {\enquote {\bibinfo {title} {Accelerators for inertial
  fusion energy production},}\ }\href {\doibase 10.1142/S1793626813300053}
  {\bibfield  {journal} {\bibinfo  {journal} {Reviews of Accelerator Science
  and Technology}\ }\textbf {\bibinfo {volume} {06}},\ \bibinfo {pages}
  {85--116} (\bibinfo {year} {2013})}\BibitemShut {NoStop}%
\bibitem [{\citenamefont {Lindemuth}\ and\ \citenamefont
  {Siemon}(2009)}]{Lindemuth}%
  \BibitemOpen
  \bibfield  {author} {\bibinfo {author} {\bibfnamefont {I.~R.}\ \bibnamefont
  {Lindemuth}}\ and\ \bibinfo {author} {\bibfnamefont {R.~E.}\ \bibnamefont
  {Siemon}},\ }\bibfield  {title} {\enquote {\bibinfo {title} {The fundamental
  parameter space of controlled thermonuclear fusion},}\ }\href {\doibase
  10.1119/1.3096646} {\bibfield  {journal} {\bibinfo  {journal} {American
  Journal of Physics}\ }\textbf {\bibinfo {volume} {77}},\ \bibinfo {pages}
  {407} (\bibinfo {year} {2009})}\BibitemShut {NoStop}%
\bibitem [{\citenamefont {Wurden}\ \emph {et~al.}(2016)\citenamefont {Wurden},
  \citenamefont {Hsu}, \citenamefont {Intrator}, \citenamefont {Grabowski},
  \citenamefont {Degnan}, \citenamefont {Domonkos}, \citenamefont {Turchi},
  \citenamefont {Campbell}, \citenamefont {Sinars}, \citenamefont {Herrmann},
  \citenamefont {Betti}, \citenamefont {Bauer}, \citenamefont {Lindemuth},
  \citenamefont {Siemon}, \citenamefont {Miller}, \citenamefont {Laberge},\
  and\ \citenamefont {Delage}}]{Wurden2016}%
  \BibitemOpen
  \bibfield  {author} {\bibinfo {author} {\bibfnamefont {G.~A.}\ \bibnamefont
  {Wurden}}, \bibinfo {author} {\bibfnamefont {S.~C.}\ \bibnamefont {Hsu}},
  \bibinfo {author} {\bibfnamefont {T.~P.}\ \bibnamefont {Intrator}}, \bibinfo
  {author} {\bibfnamefont {T.~C.}\ \bibnamefont {Grabowski}}, \bibinfo {author}
  {\bibfnamefont {J.~H.}\ \bibnamefont {Degnan}}, \bibinfo {author}
  {\bibfnamefont {M.}~\bibnamefont {Domonkos}}, \bibinfo {author}
  {\bibfnamefont {P.~J.}\ \bibnamefont {Turchi}}, \bibinfo {author}
  {\bibfnamefont {E.~M.}\ \bibnamefont {Campbell}}, \bibinfo {author}
  {\bibfnamefont {D.~B.}\ \bibnamefont {Sinars}}, \bibinfo {author}
  {\bibfnamefont {M.~C.}\ \bibnamefont {Herrmann}}, \bibinfo {author}
  {\bibfnamefont {R.}~\bibnamefont {Betti}}, \bibinfo {author} {\bibfnamefont
  {B.~S.}\ \bibnamefont {Bauer}}, \bibinfo {author} {\bibfnamefont {I.~R.}\
  \bibnamefont {Lindemuth}}, \bibinfo {author} {\bibfnamefont {R.~E.}\
  \bibnamefont {Siemon}}, \bibinfo {author} {\bibfnamefont {R.~L.}\
  \bibnamefont {Miller}}, \bibinfo {author} {\bibfnamefont {M.}~\bibnamefont
  {Laberge}}, \ and\ \bibinfo {author} {\bibfnamefont {M.}~\bibnamefont
  {Delage}},\ }\bibfield  {title} {\enquote {\bibinfo {title} {Magneto-inertial
  fusion},}\ }\href {\doibase 10.1007/s10894-015-0038-x} {\bibfield  {journal}
  {\bibinfo  {journal} {Journal of Fusion Energy}\ }\textbf {\bibinfo {volume}
  {35}},\ \bibinfo {pages} {69} (\bibinfo {year} {2016})}\BibitemShut {NoStop}%
\bibitem [{\citenamefont {Shi}\ and\ \citenamefont {Lal}(2011)}]{Shi2011}%
  \BibitemOpen
  \bibfield  {author} {\bibinfo {author} {\bibfnamefont {Y.}~\bibnamefont
  {Shi}}\ and\ \bibinfo {author} {\bibfnamefont {A.}~\bibnamefont {Lal}},\
  }\bibfield  {title} {\enquote {\bibinfo {title} {Integrated all-electric high
  energy ion beam guidance on chip: Towards miniature particle accelerator},}\
  }in\ \href {\doibase 10.1109/memsys.2011.5734380} {\emph {\bibinfo
  {booktitle} {2011 {IEEE} 24th International Conference on Micro Electro
  Mechanical Systems}}}\ (\bibinfo  {publisher} {Institute of Electrical and
  Electronics Engineers ({IEEE})},\ \bibinfo {year} {2011})\ pp.\ \bibinfo
  {pages} {137--140}\BibitemShut {NoStop}%
\bibitem [{\citenamefont {Maschke}(1979)}]{Maschke_1979}%
  \BibitemOpen
  \bibfield  {author} {\bibinfo {author} {\bibfnamefont {A.}~\bibnamefont
  {Maschke}},\ }\href {\doibase 10.2172/5914736} {\enquote {\bibinfo {title}
  {Space-charge limits for linear accelerators},}\ }\bibinfo {type} {Tech.
  Rep.}\ \bibinfo {number} {BNL-51022}\ (\bibinfo  {institution} {Brookhaven
  National Lab.},\ \bibinfo {year} {1979})\BibitemShut {NoStop}%
\bibitem [{\citenamefont {Hogan}, \citenamefont {Bangerter},\ and\
  \citenamefont {Kulcinski}(1992)}]{Hogan}%
  \BibitemOpen
  \bibfield  {author} {\bibinfo {author} {\bibfnamefont {W.~J.}\ \bibnamefont
  {Hogan}}, \bibinfo {author} {\bibfnamefont {R.}~\bibnamefont {Bangerter}}, \
  and\ \bibinfo {author} {\bibfnamefont {G.~L.}\ \bibnamefont {Kulcinski}},\
  }\bibfield  {title} {\enquote {\bibinfo {title} {Energy from inertial
  fusion},}\ }\href {\doibase 10.1063/1.881319} {\bibfield  {journal} {\bibinfo
   {journal} {Physics Today}\ }\textbf {\bibinfo {volume} {45}},\ \bibinfo
  {pages} {42--50} (\bibinfo {year} {1992})}\BibitemShut {NoStop}%
\bibitem [{\citenamefont {Faltens}\ and\ \citenamefont
  {Seidl}(1996)}]{Faltens}%
  \BibitemOpen
  \bibfield  {author} {\bibinfo {author} {\bibfnamefont {A.}~\bibnamefont
  {Faltens}}\ and\ \bibinfo {author} {\bibfnamefont {P.}~\bibnamefont
  {Seidl}},\ }\bibfield  {title} {\enquote {\bibinfo {title} {Development of
  electrostatic quadrupoles for heavy ion fusion},}\ }in\ \href {\doibase
  10.1109/DEIV.1996.545406} {\emph {\bibinfo {booktitle} {Proceedings of 17th
  International Symposium on Discharges and Electrical Insulation in
  Vacuum}}},\ Vol.~\bibinfo {volume} {1}\ (\bibinfo {year} {1996})\ pp.\
  \bibinfo {pages} {478--481}\BibitemShut {NoStop}%
\bibitem [{\citenamefont {Slade}\ and\ \citenamefont {Taylor}(2002)}]{Slade}%
  \BibitemOpen
  \bibfield  {author} {\bibinfo {author} {\bibfnamefont {P.~G.}\ \bibnamefont
  {Slade}}\ and\ \bibinfo {author} {\bibfnamefont {E.~D.}\ \bibnamefont
  {Taylor}},\ }\bibfield  {title} {\enquote {\bibinfo {title} {Electrical
  breakdown in atmospheric air between closely spaced (0.2 $\mu$m - 40 $\mu$m)
  electrical contacts},}\ }\href {\doibase 10.1109/TCAPT.2002.804615}
  {\bibfield  {journal} {\bibinfo  {journal} {IEEE Transactions on Components
  and Packaging Technologies}\ }\textbf {\bibinfo {volume} {25}},\ \bibinfo
  {pages} {390--396} (\bibinfo {year} {2002})}\BibitemShut {NoStop}%
\bibitem [{\citenamefont {Skalyga}\ \emph {et~al.}(2016)\citenamefont
  {Skalyga}, \citenamefont {Izotov}, \citenamefont {Golubev}, \citenamefont
  {Sidorov}, \citenamefont {Razin}, \citenamefont {Vodopyanov}, \citenamefont
  {Tarvainen}, \citenamefont {Koivisto},\ and\ \citenamefont
  {Kalvas}}]{Skalyga}%
  \BibitemOpen
  \bibfield  {author} {\bibinfo {author} {\bibfnamefont {V.}~\bibnamefont
  {Skalyga}}, \bibinfo {author} {\bibfnamefont {I.}~\bibnamefont {Izotov}},
  \bibinfo {author} {\bibfnamefont {S.}~\bibnamefont {Golubev}}, \bibinfo
  {author} {\bibfnamefont {A.}~\bibnamefont {Sidorov}}, \bibinfo {author}
  {\bibfnamefont {S.}~\bibnamefont {Razin}}, \bibinfo {author} {\bibfnamefont
  {A.}~\bibnamefont {Vodopyanov}}, \bibinfo {author} {\bibfnamefont
  {O.}~\bibnamefont {Tarvainen}}, \bibinfo {author} {\bibfnamefont
  {H.}~\bibnamefont {Koivisto}}, \ and\ \bibinfo {author} {\bibfnamefont
  {T.}~\bibnamefont {Kalvas}},\ }\bibfield  {title} {\enquote {\bibinfo {title}
  {New progress of high current gasdynamic ion source (invited)},}\ }\href
  {\doibase 10.1063/1.4934213} {\bibfield  {journal} {\bibinfo  {journal}
  {Review of Scientific Instruments}\ }\textbf {\bibinfo {volume} {87}},\
  \bibinfo {pages} {02A716} (\bibinfo {year} {2016})}\BibitemShut {NoStop}%
\bibitem [{\citenamefont {Ji}\ \emph {et~al.}(2016)\citenamefont {Ji},
  \citenamefont {Seidl}, \citenamefont {Waldron}, \citenamefont {Takakuwa},
  \citenamefont {Friedman}, \citenamefont {Grote}, \citenamefont {Persaud},
  \citenamefont {Barnard},\ and\ \citenamefont {Schenkel}}]{Ji16}%
  \BibitemOpen
  \bibfield  {author} {\bibinfo {author} {\bibfnamefont {Q.}~\bibnamefont
  {Ji}}, \bibinfo {author} {\bibfnamefont {P.~A.}\ \bibnamefont {Seidl}},
  \bibinfo {author} {\bibfnamefont {W.~L.}\ \bibnamefont {Waldron}}, \bibinfo
  {author} {\bibfnamefont {J.~H.}\ \bibnamefont {Takakuwa}}, \bibinfo {author}
  {\bibfnamefont {A.}~\bibnamefont {Friedman}}, \bibinfo {author}
  {\bibfnamefont {D.~P.}\ \bibnamefont {Grote}}, \bibinfo {author}
  {\bibfnamefont {A.}~\bibnamefont {Persaud}}, \bibinfo {author} {\bibfnamefont
  {J.~J.}\ \bibnamefont {Barnard}}, \ and\ \bibinfo {author} {\bibfnamefont
  {T.}~\bibnamefont {Schenkel}},\ }\bibfield  {title} {\enquote {\bibinfo
  {title} {Development and testing of a pulsed helium ion source for probing
  materials and warm dense matter studies},}\ }\href {\doibase
  10.1063/1.4932569} {\bibfield  {journal} {\bibinfo  {journal} {Review of
  Scientific Instruments}\ }\textbf {\bibinfo {volume} {87}},\ \bibinfo {pages}
  {02B707} (\bibinfo {year} {2016})}\BibitemShut {NoStop}%
\bibitem [{\citenamefont {Todd}\ \emph {et~al.}(2013)\citenamefont {Todd},
  \citenamefont {Benitez}, \citenamefont {Covo}, \citenamefont {Franzen},
  \citenamefont {Lyneis}, \citenamefont {L.~Phair},\ and\ \citenamefont
  {Strohmeier}}]{Todd-cyclotron2013}%
  \BibitemOpen
  \bibfield  {author} {\bibinfo {author} {\bibfnamefont {D.~S.}\ \bibnamefont
  {Todd}}, \bibinfo {author} {\bibfnamefont {J.~Y.}\ \bibnamefont {Benitez}},
  \bibinfo {author} {\bibfnamefont {M.~K.}\ \bibnamefont {Covo}}, \bibinfo
  {author} {\bibfnamefont {K.~Y.}\ \bibnamefont {Franzen}}, \bibinfo {author}
  {\bibfnamefont {C.~M.}\ \bibnamefont {Lyneis}}, \bibinfo {author}
  {\bibfnamefont {P.~P.}\ \bibnamefont {L.~Phair}}, \ and\ \bibinfo {author}
  {\bibfnamefont {M.~M.}\ \bibnamefont {Strohmeier}},\ }\bibfield  {title}
  {\enquote {\bibinfo {title} {High current beam extraction from the 88-inch
  cyclotron at lbnl},}\ }in\ \href
  {https://accelconf.web.cern.ch/AccelConf/CYCLOTRONS2013/papers/mo2pb02.pdf}
  {\emph {\bibinfo {booktitle} {Proceedings of Cyclotrons2013}}}\ (\bibinfo
  {year} {2013})\ pp.\ \bibinfo {pages} {19--21}\BibitemShut {NoStop}%
\bibitem [{\citenamefont {Wangler}(2008)}]{Wangler2008}%
  \BibitemOpen
  \bibfield  {author} {\bibinfo {author} {\bibfnamefont {T.}~\bibnamefont
  {Wangler}},\ }\href@noop {} {\emph {\bibinfo {title} {RF Linear
  Accelerators}}},\ Physics textbook\ (\bibinfo  {publisher} {Wiley},\ \bibinfo
  {year} {2008})\ Chap.~\bibinfo {chapter} {6}\BibitemShut {NoStop}%
\bibitem [{\citenamefont {Sakai}\ and\ \citenamefont
  {Katsumata}(1985)}]{Sakai1985}%
  \BibitemOpen
  \bibfield  {author} {\bibinfo {author} {\bibfnamefont {Y.}~\bibnamefont
  {Sakai}}\ and\ \bibinfo {author} {\bibfnamefont {I.}~\bibnamefont
  {Katsumata}},\ }\bibfield  {title} {\enquote {\bibinfo {title} {An energy
  resolution formula of a three plane grids retarding field energy analyzer},}\
  }\href {http://stacks.iop.org/1347-4065/24/i=3R/a=337} {\bibfield  {journal}
  {\bibinfo  {journal} {Japanese Journal of Applied Physics}\ }\textbf
  {\bibinfo {volume} {24}},\ \bibinfo {pages} {337--341} (\bibinfo {year}
  {1985})}\BibitemShut {NoStop}%
\bibitem [{\citenamefont {Lee}\ \emph {et~al.}(1996)\citenamefont {Lee},
  \citenamefont {Gough}, \citenamefont {Kunkel}, \citenamefont {Leung},
  \citenamefont {Perkins}, \citenamefont {Pickard}, \citenamefont {Sun},
  \citenamefont {Vujic}, \citenamefont {Williams},\ and\ \citenamefont
  {Wutte}}]{Lee1996}%
  \BibitemOpen
  \bibfield  {author} {\bibinfo {author} {\bibfnamefont {Y.}~\bibnamefont
  {Lee}}, \bibinfo {author} {\bibfnamefont {R.}~\bibnamefont {Gough}}, \bibinfo
  {author} {\bibfnamefont {W.}~\bibnamefont {Kunkel}}, \bibinfo {author}
  {\bibfnamefont {K.}~\bibnamefont {Leung}}, \bibinfo {author} {\bibfnamefont
  {L.}~\bibnamefont {Perkins}}, \bibinfo {author} {\bibfnamefont
  {D.}~\bibnamefont {Pickard}}, \bibinfo {author} {\bibfnamefont
  {L.}~\bibnamefont {Sun}}, \bibinfo {author} {\bibfnamefont {J.}~\bibnamefont
  {Vujic}}, \bibinfo {author} {\bibfnamefont {M.}~\bibnamefont {Williams}}, \
  and\ \bibinfo {author} {\bibfnamefont {D.}~\bibnamefont {Wutte}},\ }\bibfield
   {title} {\enquote {\bibinfo {title} {A compact filament-driven multicusp ion
  source},}\ }\href {\doibase http://dx.doi.org/10.1016/S0168-583X(96)00623-4}
  {\bibfield  {journal} {\bibinfo  {journal} {Nuclear Instruments and Methods
  in Physics Research Section B: Beam Interactions with Materials and Atoms}\
  }\textbf {\bibinfo {volume} {119}},\ \bibinfo {pages} {543--548} (\bibinfo
  {year} {1996})}\BibitemShut {NoStop}%
\bibitem [{\citenamefont {Kwan}, \citenamefont {Grote},\ and\ \citenamefont
  {Westenskow}(2004)}]{Kwan2004}%
  \BibitemOpen
  \bibfield  {author} {\bibinfo {author} {\bibfnamefont {J.~W.}\ \bibnamefont
  {Kwan}}, \bibinfo {author} {\bibfnamefont {D.~P.}\ \bibnamefont {Grote}}, \
  and\ \bibinfo {author} {\bibfnamefont {G.~A.}\ \bibnamefont {Westenskow}},\
  }\bibfield  {title} {\enquote {\bibinfo {title} {High current density
  beamlets from a rf argon source for heavy ion fusion applications},}\ }\href
  {\doibase 10.1063/1.1699517} {\bibfield  {journal} {\bibinfo  {journal}
  {Review of Scientific Instruments}\ }\textbf {\bibinfo {volume} {75}},\
  \bibinfo {pages} {1838--1840} (\bibinfo {year} {2004})}\BibitemShut {NoStop}%
\bibitem [{\citenamefont {Grote}, \citenamefont {Henestroza},\ and\
  \citenamefont {Kwan}(2003)}]{Grote-injector2003}%
  \BibitemOpen
  \bibfield  {author} {\bibinfo {author} {\bibfnamefont {D.~P.}\ \bibnamefont
  {Grote}}, \bibinfo {author} {\bibfnamefont {E.}~\bibnamefont {Henestroza}}, \
  and\ \bibinfo {author} {\bibfnamefont {J.~W.}\ \bibnamefont {Kwan}},\
  }\bibfield  {title} {\enquote {\bibinfo {title} {Design and simulation of a
  multibeamlet injector for a high current accelerator},}\ }\href {\doibase
  10.1103/PhysRevSTAB.6.014202} {\bibfield  {journal} {\bibinfo  {journal}
  {Phys. Rev. ST-AB}\ }\textbf {\bibinfo {volume} {6}},\ \bibinfo {pages}
  {014202} (\bibinfo {year} {2003})}\BibitemShut {NoStop}%
\bibitem [{\citenamefont {Qin}\ \emph {et~al.}(2005)\citenamefont {Qin},
  \citenamefont {Davidson}, \citenamefont {Barnard},\ and\ \citenamefont
  {Lee}}]{Qin-compression2005}%
  \BibitemOpen
  \bibfield  {author} {\bibinfo {author} {\bibfnamefont {H.}~\bibnamefont
  {Qin}}, \bibinfo {author} {\bibfnamefont {R.~C.}\ \bibnamefont {Davidson}},
  \bibinfo {author} {\bibfnamefont {J.~J.}\ \bibnamefont {Barnard}}, \ and\
  \bibinfo {author} {\bibfnamefont {E.~P.}\ \bibnamefont {Lee}},\ }\bibfield
  {title} {\enquote {\bibinfo {title} {Drift compression and final focus
  options for heavy ion fusion},}\ }\href {\doibase 10.1016/j.nima.2005.01.214}
  {\bibfield  {journal} {\bibinfo  {journal} {Nucl. Instr. Meth. A}\ }\textbf
  {\bibinfo {volume} {544}},\ \bibinfo {pages} {255--261} (\bibinfo {year}
  {2005})}\BibitemShut {NoStop}%
\bibitem [{\citenamefont {Urbanus}\ \emph {et~al.}(1990)\citenamefont
  {Urbanus}, \citenamefont {Wojke}, \citenamefont {Bannenberg}, \citenamefont
  {Klein}, \citenamefont {Schempp}, \citenamefont {Thomae}, \citenamefont
  {Weis},\ and\ \citenamefont {Van~Amersfoort}}]{Urbanus1990}%
  \BibitemOpen
  \bibfield  {author} {\bibinfo {author} {\bibfnamefont {W.~H.}\ \bibnamefont
  {Urbanus}}, \bibinfo {author} {\bibfnamefont {R.~G.~C.}\ \bibnamefont
  {Wojke}}, \bibinfo {author} {\bibfnamefont {J.~G.}\ \bibnamefont
  {Bannenberg}}, \bibinfo {author} {\bibfnamefont {H.}~\bibnamefont {Klein}},
  \bibinfo {author} {\bibfnamefont {A.}~\bibnamefont {Schempp}}, \bibinfo
  {author} {\bibfnamefont {R.~W.}\ \bibnamefont {Thomae}}, \bibinfo {author}
  {\bibfnamefont {T.}~\bibnamefont {Weis}}, \ and\ \bibinfo {author}
  {\bibfnamefont {P.~W.}\ \bibnamefont {Van~Amersfoort}},\ }\bibfield  {title}
  {\enquote {\bibinfo {title} {Comparison of two types of low-{$\beta$} {RF}
  linacs: {MEQALAC} and {RFQ}},}\ }\href {\doibase
  10.1016/0168-9002(90)90340-C} {\bibfield  {journal} {\bibinfo  {journal}
  {Nucl. Instr. Meth. A}\ }\textbf {\bibinfo {volume} {290}},\ \bibinfo {pages}
  {1--10} (\bibinfo {year} {1990})}\BibitemShut {NoStop}%
\bibitem [{\citenamefont {Ratti}\ \emph {et~al.}(2002)\citenamefont {Ratti},
  \citenamefont {Ayers}, \citenamefont {Doolittle}, \citenamefont {DiGennaro},
  \citenamefont {Gough}, \citenamefont {Hoff}, \citenamefont {Keller},
  \citenamefont {MacGill}, \citenamefont {Staples}, \citenamefont {Thomae},
  \citenamefont {Virostek}, \citenamefont {Yourd},\ and\ \citenamefont
  {Aleksandrov}}]{Ratti-SNS2002}%
  \BibitemOpen
  \bibfield  {author} {\bibinfo {author} {\bibfnamefont {A.}~\bibnamefont
  {Ratti}}, \bibinfo {author} {\bibfnamefont {J.}~\bibnamefont {Ayers}},
  \bibinfo {author} {\bibfnamefont {L.}~\bibnamefont {Doolittle}}, \bibinfo
  {author} {\bibfnamefont {R.}~\bibnamefont {DiGennaro}}, \bibinfo {author}
  {\bibfnamefont {R.~A.}\ \bibnamefont {Gough}}, \bibinfo {author}
  {\bibfnamefont {M.}~\bibnamefont {Hoff}}, \bibinfo {author} {\bibfnamefont
  {R.}~\bibnamefont {Keller}}, \bibinfo {author} {\bibfnamefont
  {R.}~\bibnamefont {MacGill}}, \bibinfo {author} {\bibfnamefont
  {J.}~\bibnamefont {Staples}}, \bibinfo {author} {\bibfnamefont
  {R.}~\bibnamefont {Thomae}}, \bibinfo {author} {\bibfnamefont
  {S.}~\bibnamefont {Virostek}}, \bibinfo {author} {\bibfnamefont
  {R.}~\bibnamefont {Yourd}}, \ and\ \bibinfo {author} {\bibfnamefont
  {A.}~\bibnamefont {Aleksandrov}},\ }\bibfield  {title} {\enquote {\bibinfo
  {title} {The sns front-end commissioning},}\ }\href
  {https://accelconf.web.cern.ch/accelconf/l02/TOC/..\%5CPAPERS\%5CTU412.PDF}
  {\bibfield  {journal} {\bibinfo  {journal} {Proceedings of LINAC2002}\ ,\
  \bibinfo {pages} {329--331}} (\bibinfo {year} {2002})}\BibitemShut {NoStop}%
\end{thebibliography}

%

\end{document}